\documentclass[twocolumn,a4paper]{article}
\usepackage{graphicx}
\usepackage{amsmath}
\usepackage{amsfonts}
\usepackage{amssymb}

\begin{document}

\title{Phase diagram of diblock polyampholyte solutions}
\author{M.\ Castelnovo\thanks{present address: Department of Chemistry and
Biochemistry, University of California Los Angeles, Los Angeles, California
90095} \ and J.\ F.\ Joanny\\
\textit{Institut Charles Sadron, 6 Rue Boussingault,}\\
\textit{67083 Strasbourg Cedex, France; Physico-Chimie, Institut Curie,}\\
\textit{11 Rue P. et M. Curie, 75231 Paris Cedex 05, France}}
\maketitle
\begin{abstract}
We discuss in this paper the phase diagram of a diblock polyampholyte solution
in the limit of high ionic strength as a function of concentration and charge
asymmetry. This system is shown to be very similar to solutions of so-called
charged-neutral diblock copolymers: at zero charge asymmetry the solution
phase separates into a polyelectrolyte complex and almost pure solvent. Above
a charge asymmetry threshold, the copolymers are soluble as finite 
size aggregates.
Scaling laws of the aggregates radius as a function of pH of the solution are
in qualitative agreement with experiments.
\end{abstract}

\section{Introduction}

Mixtures of oppositely charged polyelectrolytes have been studied quite
extensively over the last  thirty years by various experimental groups
\cite{dautzenbergPhilipp,overbeek,tsuchidaReview}. There are numerous motivations for these
studies such as the similarities with biological systems or the applications to
cosmetics and food industries. Depending on the stoichiometry of the
mixture, (the relative concentrations, the relative chain lengths and  charge densities), one observes mainly two types
of behavior: a macroscopic phase separation between the solvent and the
polymers, or a partial aggregation of the polymer chains
\cite{tsuchidaReview}. In both cases, one speaks about polyelectrolyte
complexation. The electrostatic attraction between chains of opposite signs is
clearly responsible of this  behavior. Despite this simple statement, there is
no deep understanding of the underlying mechanisms of polyelectrolyte
complexation. The introduction of the polyelectrolyte multilayers concept
almost ten years ago has revived the activity of this research area, mainly
due to  the numerous potential applications of the multilayers
\cite{decherscience}. The principle of the polyelectrolyte multilayers growth
is to adsorb sequentially polyelectrolytes of opposite signs on a charged
substrate. The adhesion between two consecutive layers is attributed to the
complexation between the oppositely charged polyelectrolytes at the interface
between layers where they strongly interpenetrate \cite{ladam,multicastel}. 

In a bulk solution, the complex formation is very similar to the demixing phase transition
of a neutral polymer solution under
poor solvent conditions if the polyelectrolyte mixture is symmetric in charge, \emph{i.e.} if the  
polycations and poyanions carry the same charge in absolute value, and their
total concentrations are the same: there is a polymer-solvent
phase separation. If the mixture is no longer symmetric in charge, 
soluble complexes carrying a net charge can exist in solution \cite{tsuchidaReview}. The aggregation process depends
on the charge asymmetry between
the polyelectrolytes, but also on the relative concentrations of the polyions.
Due to the large number of degrees of freedom, a comprehensive study of an
asymmetric polyelectrolyte mixture is quite difficult. Everaers \emph{et al.} have performed
recently a study of solutions of random
polyampholytes which are very similar systems \cite{ralph1}. They have allowed both for a macroscopic
phase separation and for a finite size aggregation process but they
have restricted themselves to some specific limits for a sake of simplicity. In
the case of a mixture of oppositely charged polyelectrolytes, one can suppress 
one degree of freedom by considering 
diblock polyampholyte solutions. These copolymers are  made of two oppositely
charged polyelectrolytes chemically linked by one end. The
concentrations of polycations and polyanions in the solution are then equal. In
particular, each aggregate is characterized by a single
aggregation number. In the equivalent polyelectrolyte mixture,
where the polyions are not bound, the aggregates are 
characterized by two aggregation numbers, the number of
polycations and polyanions respectively.

Our aim in this article is to study the phase diagram of diblock
polyampholyte solutions as a function of the copolymers charge asymmetry. This
is a first step towards a more precise understanding of aggregation
processes in mixtures of oppositely charged polyelectrolytes. Diblock
polyampholyte solutions have been far less studied experimentally than the
equivalent mixture of free polyions. Nevertheless, recent experiments 
have shown that the aggregates in
solution are roughly spherical; they adsorb  on charged substrate mostly 
as micelles, and not as isolated molecules 
\cite{stamm1,stamm2,cohendibloc}. The aggregate sizes can be measured by 
Dynamic Light Scattering experiments and are
of the same order as the size of an adsorbed aggregate. To our knowledge, in all
of the existing experiments, the charged groups on the copolymers are
pH-dependent. In particular, the copolymer has an isoelectric point at a value pH$_{i}$ of the pH,
where the net charge vanishes. Around the isoelectric point, the
solution is not stable: there is a macroscopic phase separation which is
attributed to the complexation of the copolymers. For pH values not too close
to pH$_{i}$, aggregates of diblock polyampholytes form in the
solution, but the solution is macroscopically stable. Therefore, our theoretical description must consider both possibilities of
macroscopic phase separation and of  partial aggregation of the chains in the
supernatant. This is achieved in several steps.

In a first step (section \ref{SectionThermo}) \cite{ralph1}, we neglect the influence of the supernatant content on the
properties of the dense phase for diblock polyampholytes with a quenched homogeneous charge distribution. It has been realized quite early
that the properties of polyelectrolyte complexes can be described in the low
charging limit by an equilibrium between the attraction induced by charge fluctuations and the hard core interactions \cite{overbeek}: due to the very low
mixing entropy of polymers, polymer mixtures are very sensitive  to small interactions
\cite{degennes}. This low mixing entropy is, for example, 
responsible for the so-called ``incompatibility'' between  neutral polymers. In polyelectrolyte mixtures, the phase diagram is determined by a subtle
balance between the incompatibility and complexation transitions
\cite{complexcastel}. Using our previous results on the complexation between 
oppositely charged polyelectrolytes in a symmetric mixture, we describe the
swelling of the dense phase by increase of the counterion excess osmotic pressure with the
asymmetry of the copolymers \cite{multicastel}. In a second step  (cf
section \ref{SectionMicell}), the
aggregation process is described in the absence of any macroscopic phase separation. 
By analogy with the associating behavior of
charged-neutral diblock copolymers (ionized hydrophilic block-neutral
hydrophobic block), we propose that the copolymers form 
spherical micelles with a neutral core made of symmetric 
polyelectrolyte
complex, and branches carrying the net charge of the aggregates. Finally, the
phase diagram is drawn by combining all these results in section
\ref{SectionPhasediag}. We also discuss in this section the effect of ``charge annealing'' on the properties of the aggregates in the case
where the ionizable groups along the blocks are weak acids and weak bases. All
the energies in this paper are measured in units of the thermal excitation $k_{B}T$.

\section{Complexation thermodynamics of an asymmetric polyelectrolyte mixture}

\label{SectionThermo} We consider in this section a monodisperse solution of
diblock polyampholytes of total degree of polymerization $N$. We assume that all the chemical
properties of the blocks are identical except their charge that are of opposite signs.
Let $N_{+},f_{+}$ and $N_{-},f_{-}$ be respectively the degrees of polymerization and charge
densities of the polycationic and polyanionic blocks of the copolymer. We
neglect in this paper the effects of charge distribution and assume that the
total charge of each block is smeared out along chain. We
assume that the solvent is a $\Theta$-solvent and we treat it as a dielectric continuum of dielectric constant
$\varepsilon$.  The monomer size is denoted by $a$. It is also convenient to
introduce the Bjerrum length $l_{B}=\frac{e^{2}}{4\pi\varepsilon k_{B}T}$,
which represents the typical distance between two elementary charges having an
electrostatic energy of $k_{B}T$. When the polyampholyte is symmetric in
charge, \emph{i.e.} $N_{+}f_{+}=N_{-}f_{-}$, the solution phase separates into
a dense phase containing almost all the copolymers and a dilute phase composed
mainly of solvent. Within the dense phase, the diblock copolymer  concentration (expressed as a number of chains per unit volume)
$c_{dense}$ is related to cationic and anionic monomer concentrations $c_{+}$
and $c_{-}$ by $c_{+}+c_{-}=c_{dense}N$. When $N_{+}f_{+}\neq N_{-}f_{-}$,
there is an excess charge in the dense phase, which is neutralized by the
small ions of the solution. The difference in translational entropy of these
ions between the two phases produces an excess osmotic pressure which tends to
swell the dense phase. Following the results of ref \cite{multicastel}, it is
possible to describe the effect of a charge excess in the dense phase by a
generalized two phases model. The dense phase is mainly composed of
polyampholytes, while the dilute phase is modeled as a simple
electrolyte. This amounts to neglecting the formation of finite size aggregates
in the supernatant of the solution. This approximation is discussed in
the next section. The concentrations of monovalent small ions in the dense
phase and in the dilute phase are respectively $n_{\pm}$ and $n_{\pm0}=n_{0}$.
The conditions for equilibrium between the two phases are given by the equality of
various small ions chemical potentials in the two phases, and the equality of the
osmotic pressures. Therefore, we need to compute the free energies in the two
phases to write the equilibrium conditions.

As previously mentioned, the formation of polyelectrolyte complexes in
symmetric polyelectrolyte mixtures cannot be described at the simplest 
mean field level: because of the global
electroneutrality, there is no pure coulombic term in the mean field free
energy. One has to include at least fluctuations around the electroneutral
state at the lowest level to describe the complexation. This can be done
within the RPA formalism \cite{borueErukhimovich1988,borueErukhimovich1990}.
In the case of a diblock polyampholyte solution, we will assume in a first
approximation that the RPA correction to the mean field free energy of the
dense phase is the same as that of the equivalent polyelectrolyte mixture where the two blocks are
not linked: our aim in this paper is just to describe qualitatively the
phase diagram at the level of scaling laws. Nevertheless, it seems possible to
calculate more precisely the RPA free energy of the copolymers by using the
results of reference \cite{complexcastel}. This will be the topic of a future work. 
The RPA free energy density of the dense phase is
written
\begin{eqnarray}
F_{dense} & = & \frac{w^{2}}{6}\left(  c_{+}+c_{-}\right)  ^{3}+\sum_{i}
n_{i}\log n_{i}\nonumber\\
& &  +\Delta F_{dense}\label{Fdense}
\end{eqnarray}
The first term of eq. (\ref{Fdense}) is the third virial contribution. This is
the first hard core repulsive interaction for a polymer in a $\Theta$-solvent. This 
contribution is necessary to insure the stability of the complex. The second
term is the translational entropy of small ions in the dense phase. Notice
that the translational entropy of the copolymers has been neglected because it is
of order $\frac{1}{N}$. Finally the last term is the RPA correction to the
mean field free energy. It is calculated by associating a Gaussian statistical
weight to the concentration fluctuations neglected in the mean field theory
\cite{borueErukhimovich1988}. In the case of a symmetric mixture $\left(
f_{+}c_{+}=f_{-}c_{-}\right)  $, the exact result reads
\begin{equation}
\Delta F_{dense}=-\left(  \frac{\xi_{w}^{-3}}{12\pi}+\frac{q_{\ast}^{3}\left(
s-1\right)  \left(  s+2\right)  ^{1/2}}{12\pi}\right)  \label{DeltFdense}
\end{equation}
We introduced in the last equation three characteristic lengths $\xi
_{w},q_{\ast}^{-1},\kappa^{-1}$ which are defined as follows: $\xi_{w}
^{-2}=\frac{12w^{2}\left(  c_{+}+c_{-}\right)  ^{2}}{a^{2}},$ $q_{\ast}
^{4}=\frac{48\pi l_{B}\left(  f_{+}^{2}c_{+}+f_{-}^{2}c_{-}\right)  }{a^{2}},$
$\kappa^{2}=4\pi l_{B}\left(  \sum_{i}n_{i}\right)  $. The first length is the
correlation length of the equivalent neutral polymer mixture in a $\Theta$-solvent. The two other
lengths are associated respectively to the screening of electrostatic
interactions by the polyelectrolytes and by the small ions. The dimensionless
ratio of these two lengths is denoted $s=\frac{\kappa^{2}}{q_{\ast}^{2}}$.
Notice that in the symmetrical case, the contribution of fluctuations
associated respectively to the third virial term and the electrostatics are
not coupled. The free energy density of the dilute phase is simply given
within the RPA by the Debye-H\"{u}ckel theory \cite{landau}
\begin{equation}
F_{dilute}=\sum_{i}n_{i0}\log n_{i0}-\frac{\kappa_{0}^{3}}{12\pi}
\end{equation}
with $\kappa_{0}^{2}=4\pi l_{B}\left(  \sum_{i}n_{i0}\right)  $. In the limit
of high ionic strength, one can solve the equilibrium equations using an
expansion in the small parameter $s_{0}^{-1}$. Additionally, an expansion up to 
first order in  $\kappa_{0}l_{B}$ ensures that the Debye-H\"{u}ckel
approach is valid. In the symmetric case, the osmotic balance between the two
phases reads the lowest non-trivial order
\begin{equation}
\Delta\Pi=\frac{\bar{w}^{2}}{3}\left(  c_{+}+c_{-}\right)  ^{3}-\frac
{\kappa_{0}^{3}s_{0}^{-3}}{12\pi}=0
\end{equation}
The effect of fluctuations associated to the third virial term can be absorbed
by a renormalization of $w$. The renormalized third virial term is denoted
$\bar{w}$. This osmotic balance lead us to introduce the notion of
`complexation blobs' \cite{multicastel,thesecastel}: by analogy with neutral
polymers under poor solvent conditions, the dense phase can be viewed as a
compact packing of complexation blobs of size $\xi_{c}\sim\kappa_{0}^{-1}
s_{0}$. At length scales below $\xi_{c}$, the concentration fluctuations
induced attraction is not relevant compared to the thermal energy $k_{B}T$,
while it is dominant at larger length scales. This interpretation of the
structure of the dense phase will be used in the next section to discuss the
formation aggregates.

For asymmetric mixtures $\left(  f_{+}c_{+}\neq f_{-}c_{-}\right)
$, the calculation of $\Delta F_{dense}$ is more complicated. Nevertheless we
are looking for the effect of a small charge asymmetry on the structure of the
dense phase. The calculation can be therefore simplified by expanding the RPA
correction term with the charge asymmetry \cite{thesecastel}. The main
difference with the symmetric calculation of $\Delta F_{dense}$ is the
appearance of a new term coupling the fluctuations associated to
the third virial coefficient and the electrostatic interactions. It can be shown \emph{a
posteriori }that this term simply renormalizes the prefactors in the osmotic
balance. It will be therefore neglected. The equilibrium of the small ion
chemical potentials between the two phases requires the introduction of a
Donnan potential, because of the electroneutrality constraint in the two
phases \cite{hill}. Therefore the concentrations of small ions in the dense
phase are related by
\begin{equation}
n_{+}n_{-}=n_0 ^2 \exp\left[  -2\Delta\mu_{pol}\right]
\end{equation}
with the chemical potential difference due to the polarization energy (RPA
correction term) $\Delta\mu_{pol}=\frac{\partial}{\partial n_{\pm}}\left(
\Delta F_{dense}-\Delta F_{dilute}\right)  $. This equation in $n_{+}$ can be
solved with the constraint of electroneutrality at the first order in
$\kappa_{0}l_{B}$. Finally, the osmotic balance between the two phases reads
at the lowest non-trivial order
\begin{eqnarray}
\Delta\Pi & = & \frac{w^{2}c_{dense}^{3}N^{3}}{3}+\frac{c_{dense}^{2}\left(
f_{+}N_{+}-f_{-}N_{-}\right)  ^{2}}{4n_{0}}\nonumber\\
& &  -\frac{\kappa_{0}^{3}s_{0}^{-3}}{24\pi} \label{DelPi}
\end{eqnarray}
We used in the preceding equation the definition of the monomer concentrations in
the dense phase $c_{\pm}=c_{dense}N_{\pm}$. The first and third terms of this
equation are reminiscent of the osmotic pressure of a symmetrical complex. The second term is of purely
entropic origin: the difference in translational entropy of the small ions between
the two phases induces an excess osmotic pressure inside the dense phase. The
excess charge per copolymer is conveniently measured by the introduction of
the \emph{average net charge }per monomers
\begin{equation}
q\equiv\frac{f_{+}N_{+}-f_{-}N_{-}}{N}
\end{equation}
This is the net charge of a copolymer redistributed on all the monomers. In the
limit of low asymmetry, the third virial term dominates over the osmotic term.
The copolymer concentration inside the dense phase is therefore given by its
value in the symmetrical case
\begin{equation}
c_{dense}a^{3}N=\frac{3}{2\pi^{2/3}}\frac{f_{+}f_{-}}{n_{0}a^{3}\left(
\frac{w^{1/3}}{a}\right)  ^{4}} \label{cdense}
\end{equation}
The number of monomers inside a complexation blob reads
\begin{equation}
g_{c}\sim\frac{\left(  n_{0}a^{3}\right)  ^{2}\left(  \frac{w^{1/3}}
{a}\right)  ^{2}}{\left(  f_{+}f_{-}\right)  ^{2}} \label{gc}
\end{equation}
The size of the complexation blob is given by $\xi_{c}\sim g_{c}^{1/2}a(w^{1/3}/a)$. Neglecting all
numerical prefactors, a careful analysis of eq. (\ref{DelPi}) shows that the
third virial contribution dominates over the osmotic one if
\begin{equation}
q<\left(  f_{+}f_{-}\right)  ^{1/2}\left(  \frac{w^{1/3}}{a}\right)
\end{equation}
This inequality is valid as long as $\arrowvert f_{+}-f_{-}\arrowvert\frac{w^{1/3}}{a}\lesssim (f_+ f_-)^{1/2}$. It means that the properties of the dense phase are well described by
equations (\ref{cdense}) and (\ref{gc}) if $\frac{q}{f}\lesssim1$, where $f$
is the geometric average of charge densities. If $\frac{q}{f}\gtrsim1$, the
osmotic term dominates over the third virial term, and the copolymer
concentration of the dense phase can be calculated using the osmotic balance
eq. (\ref{DelPi}). However, as will be shown in the next sections, finite size
aggregates are more favorable in this range of parameters.

\section{Micelle formation}

\label{SectionMicell}
\begin{figure}
[h]
\begin{center}
\includegraphics[scale=0.5]
{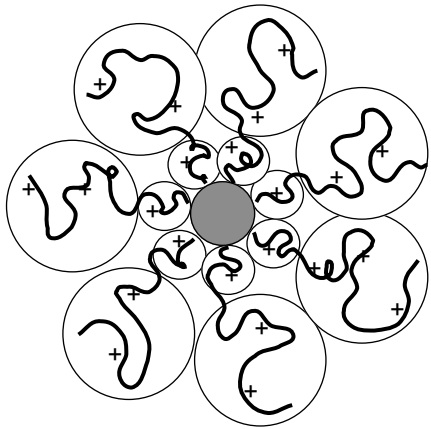}
\caption{Copolymer micelle. The circles of growing diameters are the compact
packing blobs of the Daoud Cotton model. For the sake of simplicity, only
charges of the arms are explicitly sketched.}
\label{FigDc}
\end{center}
\end{figure}

We propose in this section a model for the aggregates in
the supernatant. It has been shown in the last section that the polymeric net
charge in the dense phase induces an excess osmotic pressure which tends to
swell this phase. By analogy with the associating
behavior of diblock copolymers charged hydrophilic/hydrophobic, we will assume
that spherical micelles are formed in solution \cite{rolanddibloc}: the
polymeric net charge of the core vanishes, and the excess charge is localized
on the surrounding arms in the corona of the micelle (cf figure \ref{FigDc}).
For a positive excess charge per copolymer, the number of cationic monomers in
the core $N_{c+}$ is given by the electroneutrality condition $N_{c+}
f_{+}=N_{-}f_{-}$. The total number of monomers per chain in the core is
denoted by $N_{c}=N_{c+}+N_{-}$. The excess charge of the corona per copolymer is
$Nq$. In this geometry, the non-compensated charges of the arms are
further away from each other than in a single homogeneous dense phase. It is
also the case for charged-neutral diblock copolymers where this geometry has
been observed experimentally \cite{ExempleDiblocNC}. When the charge asymmetry
is small enough, we expect other aggregate geometries to be more favorable
than the spherical one, such as cylindrical micelles or lamellae (planar
geometry) \cite{carlosdibloc}. However it can be shown \emph{a posteriori}
that those geometries are less favorable at the level of scaling laws than a
macroscopic phase separation for low charge asymmetry \cite{thesecastel}.
Therefore we will not consider explicitly those geometries further.
Before drawing the full phase diagram of the solution, we first consider in this section, the
properties of the optimal spherical micelle, which minimizes the free energy
per copolymer. We suppose therefore that all the required
conditions for the formation of spherical micelles are met. The main contributions
to the micelle free energy per copolymer are the surface tension of the core
and the electrostatic free energy of the corona. In the range where spherical micelles are
favorable, the stretching energy of the chains in the core is negligible
\cite{pierreDibloc}.

The ``neutral'' core of the micelle can be viewed as a symmetrical
polyelectrolyte complex. Using the interpretation of its structure in terms of
complexation blobs in the limit of high ionic strength, its negative free energy is
evaluated with the \emph{ansatz} `$k_{B}T$ per blob'. Similarly, the
interfacial tension between the complex and the surrounding solvent is estimated by
associating an energy $k_{B}T$ per blob on the surface of the core.
Therefore the surface tension reads $\gamma a^{2}\sim\frac{k_{B}Ta^{2}}
{\xi_{c}^{2}}\sim\frac{k_{B}T}{g_{c}}$. For an aggregate containing $p$ chains, the
core size $R_{c}$ is given by $R_{c}^{3}\sim pN_{c}g_{c}^{1/2}a^{3}$. The
interfacial free energy per copolymer reads
\begin{equation}
F_{surf}\sim\frac{\gamma a^{3}N_{c}g_{c}^{1/2}}{R_{c}}\sim\frac{aN_{c}}
{R_{c}g_{c}^{1/2}}
\end{equation}

In the limit of high ionic strength $n_{0}a^{3}>\frac{f_{+}^{4/3}}{\left(
l_{B}/a\right)  ^{1/3}}$, it has been shown that the corona of a charged
micelle can be described by the Daoud-Cotton model for neutral star branched
polymers \cite{borisovstarannealed}. This comes from the fact that in the high
ionic strength limit, the electrostatic interactions are strongly screened.
This effective short range interaction is equivalent to an excluded volume
interaction with an electrostatic excluded volume parameter $v_{el}\sim
\frac{f_{+}^{2}}{n_{0}}\sim\bar{v}_{el}a^{3}$. According to the Daoud-Cotton
model, the star is composed of a compact packing of correlation blobs of size
$\xi\left(  r\right)  $, where $r\,$\ is the distance to the center of the
star \cite{DC}. This condition implies $\xi\left(  r\right)  \sim\frac
{r}{p^{1/2}}$. The chain statistics of inside a blob the
excluded volume statistics $\xi\left(  r\right)  \sim g\left(  r\right)
^{3/5}v_{el}^{1/5}a^{2/5}$. The monomeric concentration profile of the corona
is then given by $c\left(  r\right)  a^{3}\sim\frac{p^{2/3}}{\bar{v}
_{el}^{1/3}\left(  r/a\right)  ^{4/3}}$. The extension of the arms and the
free energy per chain are calculated by counting the number of blobs, or
equivalently by
\begin{eqnarray}
\left(  N_{+}-N_{c+}\right)  p  & = & \int_{R_{c}}^{R}dr\text{ }4\pi
r^{2}c\left(  r\right) \\
F_{corona}p  & = &\int_{R_{c}}^{R}dr\frac{\text{ }4\pi r^{2}}{\xi^{3}\left(
r\right)  }
\end{eqnarray}
Let $N_{b}=N_{+}-N_{c_{+}}$ be the number of monomers per branch in 
the corona. In the limit $R/R_{c}>>1$ where the spherical aggregate is 
the most favorable geometry, one finds for the corona extension the same result as in a Flory-like mean
field evaluation balancing the osmotic pressure due to the salt in 
the micelle and
the stretching of the arms $R\sim aN_{b}^{3/5}p^{1/5}\bar{v}_{el}^{1/5}$. 
The Daoud-Cotton model takes into account excluded volume correlations, and
therefore leads to a more accurate result for the free energy than the mean-field prediction. 
The free energy per copolymer reads
\begin{equation}
F_{corona}\sim p^{1/2}\log\frac{N_{b}^{3/5}p^{1/5}\bar{v}_{el}^{1/5}a}{R_{c}}
\end{equation}
It depends only very weakly on $N_{b}$ (logarithmically).

The minimization of the total free energy per copolymer leads to the 
following radius,
aggregation number and free energy for the spherical micelle
\begin{eqnarray}
R & \sim  & a\frac{N_{b}^{3/5}N_{c}^{4/25}\bar{v}_{el}^{1/5}}{g_{c}
^{4/25}\left(  \log\frac{R}{R_{c}}\right)  ^{6/25}}\label{radius}\\
\bar{p} & \sim & \frac{N_{c}^{4/5}}{g_{c}^{4/5}\left(  \log\frac{R}{R_{c}
}\right)  ^{6/5}}\\
F_{\bar{p}} & \sim & \left(  \frac{N_{c}}{g_{c}}\right)  ^{2/5}\left(  \log
\frac{R}{R_{c}}\right)  ^{2/5}
\end{eqnarray}
It can be shown that the spherical geometry is favored if one ignores macroscopic phase separation
if $\frac{q}{f_{+}}>>\frac{N_{c}^{11/15}
g_{c}^{1/10}}{N\bar{v}_{el}^{1/3}}$. This condition is also equivalent to
$R/R_{c}>>1$.

\section{Phase diagram}

\label{SectionPhasediag}Our aim in this section is to link the the results of the two previous sections: in section \ref{SectionThermo}, we have 
studied the
macroscopic phase separation of the solution by assuming a macroscopic phase separation ignoring 
the finite size aggregation in the supernatant; in section \ref{SectionMicell}, we have evaluated the optimal properties of a spherical aggregate, 
neglecting any 
concentration effect. In fact, a diblock polyampholyte solution 
should be fully described by a model allowing both for macroscopic and mesoscopic phase
separations. Let
$\phi,c_{dense},n_{+}$ and $n_{-}$ be respectively the volume fraction, the
copolymer concentration, the positive and negative small ion concentrations in 
the dense phase of a macroscopically separated solution. In the dilute phase, we define respectively $c_{p},n_{+0}$
and $n_{-0}$ the concentrations of $p$ chains aggregates, of positive and
negative small ion. The free energy density of the solution reads
\begin{eqnarray}
F & = & \phi F_{dense}\left(  c_{dense},n_{+},n_{-}\right)  \nonumber\\
&  & +\left(  1-\phi\right)  F_{dil}\left(  \left\{  c_{p}\right\}
,n_{+0},n_{-0}\right)
\end{eqnarray}
with the free energy density of the dense phase
\begin{eqnarray}
F_{dense} & = & c_{dense}\left(  \log c_{dense}-1\right)  \nonumber\\
& &  +\sum_{i=\pm}n_{i}\left(  \log n_{i}-1 \right) +F_{D}
\end{eqnarray}
and the free energy density of the dilute phase
\begin{eqnarray}
F_{dil} & = & \sum_{p=1,2\ldots}c_{p}\left(  \log 
c_{p}-1+F_{p}\right)  \nonumber\\
& &+\sum_{i=\pm}n_{i0}\left(  \log n_{i0}-1\right)+F_{d0}
\end{eqnarray}
The quantities $F_{D},F_{p}$ and $F_{d0}$ are respectively the free
energy of  the dense phase, and the free energies of an aggregate of $p$ chains and the Debye H\"{u}ckel polarization energy
of the dilute phase. Within the RPA, $F_{D}$ includes the excluded volume contribution and the polarization energy of the dense phase eq.
(\ref{DeltFdense}), and depends on the various concentrations. Finally, $F_{p}$ has been evaluated in section
\ref{SectionMicell}. We will keep these very general notations in the
following, since they are not model dependent. The equilibrium between the two
phases is obtained by minimizing the free energy density with respect to the
volume of the dense phase and the various concentrations, with the
constraints of mass conservation of copolymers and small ions, and
electroneutrality of both phases. The minimization with respect to the volume
of the dense phase leads to the osmotic balance. When the contribution of
translational entropy and polarization energy of the aggregates are
neglected, this leads to the analysis of section \ref{SectionThermo}. This
approximation is justified since the translational entropy of an aggregate of
$p$ chains is of order $\frac{1}{pN}$, and therefore the contribution to the
polarization energy is also negligible. The other equilibrium equations read
\begin{eqnarray}
n_{+} & = & n_{+0}\exp\left(  -\Psi\right)  \exp\left(  -\left(  \mu_{i}
-\mu_{i0}\right)  \right)  \nonumber\\
n_{-} & = & n_{-0}\exp\left(  +\Psi\right)  \exp\left(  -\left(  \mu_{i}
-\mu_{i0}\right)  \right)  \label{cpVScdense}\\
c_{p} & = & c_{dense}^{p}\exp\left(  +pNq\Psi\right)  \exp\left(  -\left(
F_{p}-p\mu_{D}\right)  \right)  \nonumber
\end{eqnarray}
The chemical potentials $\mu_{\alpha}$ in the last equations are obtained by derivation of the free energy densities with respect to the 
corresponding concentration; they do not include the contributions of the translational entropies; the subscript $_0$ corresponds to the dilute phase. The
electroneutrality constraint in the two phases induce an
electrostatic potential difference between the two phases $\Psi$: this is the so-called
Donnan potential. By using the first two equations of (\ref{cpVScdense}), we
can calculate $\Psi$ at first order in $\kappa_{0}l_{B}$
\begin{equation}
\Psi=\frac{c_{dense}Nq}{2n_{0}}+o\left(  \left(  \frac{c_{dense}Nq}{2n_{0}
}\right)  ^{2}\right)  +o\left(  \left(  \kappa_{0}l_{B}\right)  ^{2}\right)
\label{PotDonnan}
\end{equation}
The last equation of (\ref{cpVScdense}) can be rewritten $c_{p}=c_{dense}
\exp\left(  -\Omega_{p}\right)  $ with the thermodynamic grand potential
\begin{equation}
\Omega_{p}=F_{p}-p\left(  \mu_{D}+\log c_{dense}+Nq\Psi\right)
\end{equation}
The total copolymer concentration $c^{tot}$ is written as
\begin{equation}
\phi c_{dense}+\left(  1-\phi\right)  (
{\textstyle\sum_{p}}
pc_{dense}\exp\left(  -\Omega_{p}\right)  )=c^{tot}
\end{equation}
This equation shows that the contribution of aggregates is negligible compared
to the contribution of the dense phase if $\Omega_{p}>0$. In the opposite case
$\Omega_{p}<0$, the concentration of aggregates becomes relevant. Therefore we
can write a simple solubility criterion: the chains are soluble as aggregates
when
\begin{equation}
\frac{F_{\bar{p}}}{\bar{p}}<\mu_{D}+\frac{c_{dense}\left(  Nq\right)  ^{2}
}{2n_{0}}+\log c_{dense}\label{critSolu}
\end{equation}
We used in this equality the average aggregation number $p=\bar{p}$ that minimizes the free
energy per diblock copolymer. This inequality means
that copolymers are soluble when the chain chemical potential in an aggregate
is smaller than the chain chemical potential in the dense phase. The dense phase
chemical potential has two main contributions: the attractive
energy, $-k_{B}T$ per blob, and the energy associated to the loss of
translational entropy of the small ions. A careful analysis of the inequality
(\ref{critSolu}) leads to the simplified criterion
\begin{equation}
\frac{N_{b}}{g_{c}}<\frac{c_{dense}\left(  Nq\right)  ^{2}}{2n_{0}}
\end{equation}
Therefore the chains are soluble when
\begin{equation}
q>f_{-}\left( \frac{w^{1/3}}{a}\right) ^{2}\label{criter}
\end{equation}
(Note that the prefactor on the right hand side must be smaller than one for consistency).
At the level of the scaling laws, we found that the dense phase disappears
when $\frac{q}{f}\gtrsim1$, where $f$ is the average charge density of the
copolymer. The chains are thus soluble as spherical aggregates when the excess
charge per monomer is of the same order of the charge density that induces the
collapse of the complex by fluctuations. This value corresponds at
the level of scaling laws to the regime where the excess osmotic pressure due
to the small ions becomes of the order of the third virial contribution in the
analysis of the osmotic balance in section \ref{SectionThermo}. The criterion
(\ref{criter}) does not depend on concentration effects. This means that the
solution has to be already phase separated or partially aggregated for this
criterion to hold. It happens only above some critical copolymer concentration.

To include concentration effects in this description, one has to consider the
two following asymptotic limits: a micellization with no macroscopic phase
separation, and a macroscopic phase separation with no micellization. This
allows to compute respectively the critical micellar concentration (CMC) and
the critical macroscopic phase separation concentration, and to compare them.
In the case of a pure micellization, the total concentration of copolymers is
written as
\begin{equation}
c_{1}+\left(  \sum\nolimits_{p}pc_{p}\right)  =c^{tot}
\end{equation}
The equilibrium conditions between all aggregates can be written as
\begin{equation}
c_{p}=c_{1}^{p}\exp\left[  -\left(  F_{p}-pF_{1}\right)  \right]
\end{equation}
where $F_{1}$ is the free energy of one isolated copolymer in solution. This
energy contains mainly two contributions: the interfacial tension between the
complexed part and the solvent $\left(  \frac{N_{c}}{g_{c}}\right)  ^{2/3}$,
and the attractive energy $-\frac{N_{c}}{g_{c}}$. Up to entropic effects,
micelles are formed in solution if $F_{1}>\frac{F_{\bar{p}}}{\bar{p}}$, where
$p=\bar{p}$ minimizes the free energy per copolymer in an aggregate. The CMC
is then evaluated by
\begin{equation}
\log c_{mc}\sim-\left(  F_{1}-\frac{F_{\bar{p}}}{\bar{p}}\right)  \label{CMC}
\end{equation}
For spherical micelles, this concentration can be rewritten as
\begin{equation}
\log c_{mc}\sim-\left(  \left(  \frac{N_{c}}{g_{v_{el}}}\right)
^{2/3}-\left(  \frac{N_{c}}{g_{v_{el}}}\right)  ^{2/5}\right)
\end{equation}
This CMC is exponentially small, as for any diblock
copolymer system.

In the limit of a pure macroscopic phase separation, with no micellization,
the total copolymer concentration reads
\begin{equation}
\phi c_{dense}+\left(  1-\phi\right)  c_{1}=c^{tot}
\end{equation}
The equality of diblock copolymer chemical potentials between the two phases leads to
\begin{equation}
c_{1}=c_{dense}\exp\left[  -\left(  F_{1}-\left(  \mu_{D}+Nq\Psi\right)
\right)  \right]
\end{equation}
We introduced in the last equation the Donnan potential that takes into
account the loss of translational entropy of small ion in the dense phase. The
critical concentration for macroscopic phase separation is thus given by
\begin{equation}
\log c_{sep}\sim-\left(  F_{1}-\left(  \left(  \mu_{D}+Nq\Psi\right)  \right)
\right)
\end{equation}
This critical concentration is approximately of the same order of the CMC.
More precisely, the ratio between those two critical concentrations reads
\begin{equation}
\log\frac{c_{mc}}{c_{sep}}\sim\frac{F_{\bar{p}}}{\bar{p}}-\left(  \mu
_{D}+Nq\Psi\right)
\end{equation}
The right hand side of the preceding equation gives our simplified criterion
for the solubility of aggregates.
\begin{figure}
[th]
\begin{center}
\includegraphics[scale=0.5]
{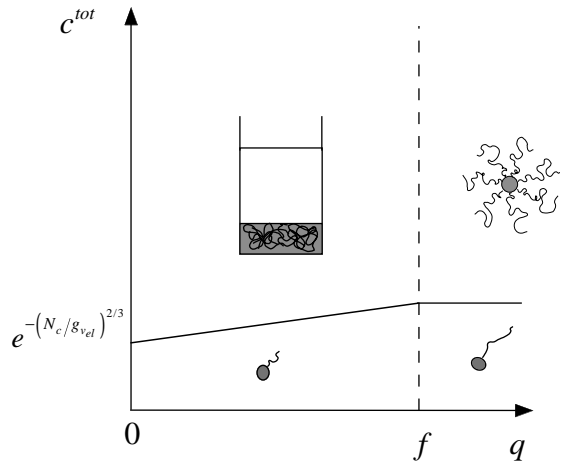}
\caption{Phase diagram ($c^{tot}$ versus $q$) in the limit of high ionic
strength.}
\label{FigDiagpad}
\end{center}
\end{figure}
Therefore, we can draw qualitatively the following phase diagram (cf figure
\ref{FigDiagpad}): the macroscopic phase separation takes place first when
$\frac{q}{f}\lesssim1$ and $c^{tot}>c_{sep}$. For higher asymmetries $\frac
{q}{f}\gtrsim1$, spherical micelles are formed when $c^{tot}>c_{mc}$. We thus
predict that the only aggregates to be observable when the solution is
macroscopically stable are spherical micelles. This has to be compared to the
recent experiments on diblock polyampholyte solutions
\cite{stamm1,stamm2,stamm3,cohendibloc}. As was mentioned in the
introduction, one observes indeed in those experiments spherical
aggregates in solution: their radius is measured by Dynamic Light
Scattering or by AFM experiments, since those copolymers mainly adsorb on
charged substrates as micelles and not as single molecules.

The only available experiments on diblock polyampholytes to our knowledge are
performed with weak acidic and basic groups on the two blocks. It means therefore
that the excess charge $q$ depends on the local pH of the solution 
\cite{distribcastel}. In
particular, the radius of the aggregates varies with the pH. As we shall see below in an appendix,
our model predicts qualitatively the same variation if local pH effects are
neglected. By local pH effects, we mean inhomogeneities of the solution (such as 
aggregation) that lead to inhomogeneities of the pH that can take locally a value different than the imposed value. For example, the pH
inside an isolated star-branched weak polyelectrolyte is different from the
value in the bulk solution \cite{borisovstarannealed}. Neglecting local pH
effects, we assume in a first approximation, that pH variations in a solution of
diblock polyampholytes change only the value of $q.$ Notice that in the most
general case, $N_{c}$ and $N_{b}$ depend in a complex manner on $q.$ It is
possible to simplify those expressions for two limiting cases of asymmetry
$\left(  N_{+}=N\text{ and }f_{+}\neq f_{-}\right)  $ and $\left(  N_{+}\neq
N_{-}\text{ and }f_{+}=f_{-}\right)  $. We will not analyze those 
particular cases in this paper. For the sake of simplicity, we
consider $N_{c}$ and $N_{b}$ as constants. The charge densities along the two
blocks are given by the following action mass laws
\begin{eqnarray}
\frac{f_{-}}{1-f_{-}} & = & \frac{K_{A}}{c_{H^{+}}}\label{acide}\\
\frac{f_{+}}{1-f_{+}} & = & 
\frac{K_{B}}{c_{OH^{-}}}=K_{B}10^{14}c_{H^{+}}\label{base}
\end{eqnarray}
with the dissociation constants $K_{A}$ and $K_{B}$ of the acidic and basic groups.
The pH of the solution is related to the concentration of $H^{+}$
ions by $c_{H^{+}}=10^{-pH}$. In eq. (\ref{base}), the ionization product of
water has been used. For weakly charged polyelectrolytes $\left(  f_{+}
,f_{-}<<1\right)  $, the product of charge densities $\ f_{+}f_{-}$ is
approximately constant and does not depend strongly on the pH. As it was shown
in the first section of this paper, the properties of the dense phase (in
macroscopically phase separated sample or in the core of a micelle) depend
only on $f_{+}f_{-}$ (for example $g_{c}\sim\left(  f_{+}f_{-}\right)  ^{-2}
$). Therefore pH variations do no affect strongly the core properties of spherical
micelles within this very crude model. The only effect of pH variations is to
change the corona extension which depends strongly on the charge density of
the arms. For a positive excess charge (low pH values), the radius of a
micelle vary as
\begin{equation}
R\sim c_{H^{+}}^{2/5}
\end{equation}
The radius of a spherical aggregate decreases with the pH. On the contrary,
the radius for a negatively charged micelle (high pH values) scales like
\begin{equation}
R\sim c_{H^{+}}^{-2/5}
\end{equation}
Therefore the extension of the micelle increases with the pH. For intermediate
values close to the isoelectric point  pH$_{i}$, the excess charge vanishes and the solution
macroscopically phase separates. Notice that eq. (\ref{radius}) is no longer
valid when $R\sim R_{c}$ since the logarithmic factor becomes important. In this range
of parameters, the solution has already phase separated as it discussed at
the beginning of this section. Close to this regime the aggregation number
increases again with the pH for pH$<$pH$_{i}$ and decreases for pH$>$pH$_{i}$.

The qualitative behavior of the micelle radius with pH is
qualitatively observed in experiments of reference \cite{stamm1}. The
authors study the adsorption of dilute diblock polyampholyte
solutions. As it previously mentioned, AFM experiments on the substrates
and Dynamic Light scattering experiments in solution confirm the presence of
spherical aggregates. By studying the adsorption kinetics as a function of the ionic strength 
of the solution, a
diffusion coefficient towards the surface is measured and can be used as an alternative mean to determine
the size of the aggregates. This radius is found to decrease with
added salt, in qualitative agreement with our model which predicts $R\sim
n_{0}^{-13/25}$. Notice that no significant variations of the aggregate radius
with pH or ionic strength is observed in the experiments of Goloub \emph{et
al. }with another kind of diblock polyampholytes. We believe that this is
probably due to the relevance of non electrostatic interactions in this
case \cite{cohendibloc}. It seems thus that
diblock polyampholyte aggregates can be described by a very crude model of
polyampholytes with quenched charge distribution with no local pH effects. The agreement 
with the experiments is not quantitative, but all
the qualitative trends are predicted. We give some arguments in the
appendix of this paper that justify partially this very crude model.

\section{Concluding remarks}

\label{SectionConclu}We have studied in this paper the phase diagram of
monodisperse diblock polyampholyte solutions.  In the case of charge symmetry $\left(  N_{+}
f_{+}=N_{-}f_{-}\right)  $, the properties of the solution
are very similar to those of an equivalent homopolymer mixture obtained by cutting the junction point 
between  the oppositely
charged polyelectrolytes in the copolymer : there is a macroscopic complexation phase separation between
the polymers and the solvent. The structure of the dense phase has  been here described in terms of
complexation blobs. When the charge asymmetry increases,
the small ions of the solution are strongly coupled to the copolymers: the
electroneutrality of the dense phase requires a finite density of counterions that induces an excess osmotic
pressure which tends to swell the complex. Above a charge asymmetry threshold,
this excess osmotic pressure is larger than the attractive energy
collapsing the complex: the copolymers are soluble in the form of finite size aggregates.
We propose in this paper that the chains are aggregated in solution as spherical
micelles with a neutral core, the excess charge being carried  by the arms
of the corona. In the limit of high ionic strength, the micelle corona
is described by the Daoud Cotton model for neutral polymeric stars with an effective electrostatic
excluded volume $v_{el}=\frac{4\pi l_{B}f^{2}}{\kappa_{0}^{2}}$. The structure
of the micelle core is very similar to a neutral polymer solution in a poor solvent
and is described using complexation blobs.

The only available experiments that we are aware of on diblock polyampholyte solutions are performed with annealed 
polyampholytes carrying acidic and
basic groups along each blocks. Therefore the charge distribution along the
chain is annealed and depend on the local pH of the solution. By neglecting
any local pH effects, it is possible to predict at least qualitatively the
variation of the micelle radius with pH and salt amount. In the limit of high
ionic strength, this simple model can be partially justified: the charge
fluctuations still collapse the core of the micelles and are in fact stronger 
in the case of an annealed charge distribution. Moreover the pH
variations of the buffer are transmitted to the corona and the core of the
micelles. Of course this very simple model is not expected to be a
quantitative description of the aggregates because of the naive modelization
of the annealed effect. The main problem for a more precise modelization comes
from the dependencies of some parameters of the  complex as a function of the charge
asymmetry: the number of monomers in the core and in the corona of the
micelle, the charge densities inside the core and on the arms of the corona,etc...

The micellization model proposed in this paper is a first step towards a more
precise understanding of asymmetries effects in mixtures of (unbound) 
oppositely
charged polyelectrolytes. In this case, the separation of the polyions produces a
new degree of freedom. In particular, the aggregates are characterized by two
aggregation numbers. This will the topic of future work.

\textbf{Acknowledgements}
\  Fruitful discussions with H.\ Mohrbach and F.\ Clement (Institut Charles
Sadron, Strasbourg) are gratefully acknowledged. This research was supported
by the Deutsche Forschung Gemeinschaft through the Schwerpunkt
program ``Polyelektrolyte''.

\section*{Appendix A}

We give briefly in this appendix some arguments justifying the very crude
model of charge annealing in the micelles in the limit of high ionic strength.

\subsection*{ A.1\ \ RPA free energy}

To understand the effect of an annealed charge distribution, let us see how the polarization
energy  driving the complexation is modified. We consider
Gaussian chains for the sake of simplicity (no third virial contribution). The
general form of the RPA correction term to the free energy is given by
\begin{equation}
\Delta F=\frac{1}{4\pi^{2}}\int_{0}^{\infty}dq\text{ }q^{2}\log\left[
\det\left[  \frac{\hat{G}_{RPA}^{-1}}{2\pi V}\right]  \right]
\end{equation}
where $\hat{G}_{RPA}^{-1}$ is the inverse structure matrix matrix at the
RPA level \cite{borueErukhimovich1988}. The determinant of this matrix in the
case of an annealed charge distribution is exactly the same as in the quenched
distribution case if the Debye-H\"{u}ckel screening length is defined as
\begin{equation}
\kappa^{2}=4\pi l_{B}\left(  \sum_{i}n_{i}+f_{+}\left(  1-f_{+}\right)
c_{+}+f_{-}\left(  1-f_{-}\right)  c_{-}\right)
\end{equation}
The mobile charges on the chains contribute to the screening of 
electrostatic interactions. The quantities $n_{i}$ are the 
concentrations of small ions (salt ions,
H$^{+},$OH$^{-}$). Notice that we will sometimes denote the concentrations of
H$^{+}$ and OH$^{-}$ by $n_{H^{+}}=c_{H^{+}}$ and $n_{OH^{-}}=c_{OH^{-}}$.
Therefore the RPA correction term reads
\begin{equation}
\Delta F=-\frac{q_{\ast}^{3}}{12\pi}\left(  s-1\right)  \left(  s+2\right)
^{1/2}
\end{equation}
with $q_{\ast}^{4}=\frac{48\pi l_{B}\left(  f_{+}^{2}c_{+}+f_{-}^{2}%
c_{-}\right)  }{a^{2}}$ and $s=\frac{\kappa^{2}}{q_{\ast}^{2}}$. It is exactly
the same as for quenched distributions, except for the new definition
of $\kappa$. The mean field contribution to the free energy density is given
by
\begin{eqnarray}
\frac{F}{kT} &  = & \frac{c_{+}+c_{-}}{N}(\log\frac{c_{+}+c_{-}}{N}
-1) \nonumber \\
& & +\sum\limits_{i}n_{i}(\log n_{i}-1) \nonumber\\
& &  +c_{-}[f_{-}\log f_{-}+(1-f_{-})\log(1-f_{-}) \nonumber\\
& & +f_{-}\mu_{A^{-}}+(1-f_{-})\mu_{AH}] \nonumber\\
& &  +c_{+}[f_{+}\log f_{+}+(1-f_{+})\log(1-f_{+}) \nonumber \\
& & +f_{+}\mu_{B^{+}}+(1-f_{+})\mu_{BOH}] \\
\end{eqnarray}
where we take into account the translational entropies of the chains and the small ions
in the solution, the mixing entropy of the small ions on the chains and their
equilibrium with a charge reservoir.

\subsection*{ A.2 \ \ Equilibrium between two phases}

As in section \ref{SectionThermo}, we neglect the influence of copolymers
in the dilute phase. We can write the equilibrium conditions as in
section \ref{SectionThermo}. In the limit of high ionic strength, one can
solve at the first order in $\kappa_{0}l_{B}$ the equilibrium equations for
the small ions concentrations \cite{thesecastel}. This leads to introduce a
new parameter $\eta=f_{+}\left(  1-f_{+}\right)  c_{+}+f_{-}\left(
1-f_{-}\right)  c_{-}$. In the symmetric case where $f_{+}c_{+}=f_{-}c_{-},$
the osmotic balance reads at lowest order in $\kappa_{0}l_{B},s_{0}^{-1}$
and$\frac{\eta}{n_{0}}$
\begin{equation}
\Delta\Pi=-\frac{\kappa_{0}^{3}}{24\pi}(s_{0}^{-3}+\frac{3}{2}\frac{\eta
}{2n_{0}}s_{0}^{-2})
\end{equation}
Thus we find that at lowest order that charge annealing leads to an
increased attractive term free energy the complex: this is quite natural since there
is a new fluctuating variable.
Nevertheless, the dominant term is still the same as in the quenched case.

\subsection*{ A.3 \ \ Local pH effects}

We denote in this section with a ``0'' subscript the quantities 
related to the region outside the micelles, namely the bulk solvent properties.
In particular, the charge density of an isolated weak polyelectrolyte (no
local pH effects) is $f_{0}$. In the case of star-branched weak
polyelectrolytes, it was shown that the charge density of the arms depends on the pH
of the buffer solution like \cite{borisovstarannealed}
\begin{equation}
f^{2}=\frac{f_{0}}{c}\left(  c_{H^{+}0}+n_{sel}+n_{strong\text{ }base}\right)
\end{equation}
if  $f_{0-}c>>\left(  c_{H^{+}0}+n_{sel}+n_{strong\text{ 
}base}\right)$.Notice that the contribution of the polarization 
energy has been neglected. In
the opposite case, the charge density is $f=f_{0}$. It means therefore
that in the high ionic strength limit, the pH variations in the solution buffer
are \emph{entirely} transmitted to the inside of the star.

All the arguments presented above can still be used: in the limit of high ionic
strength, local pH effects are very weak, and therefore they can be ignored in
first step of modelization.

\end{document}